\documentclass[twocolumn,aps,prc,floatfix]{revtex4}
\usepackage{graphicx}
\usepackage{dcolumn}
\usepackage{bm}
\usepackage{mhchem}

\begin{document}

\title{Probing in-medium nucleon-nucleon inelastic scattering cross section \\
by using energetic n/p ratio}
\author{Zong-Zhen Zhang$^{1,2}$}
\author{Ya-Fei Guo$^{1,2,3}$}
\author{Gao-Chan Yong$^{1,2}$}
\email[]{yonggaochan@impcas.ac.cn}
\affiliation{$^1$Institute of Modern Physics, Chinese Academy of Sciences, Lanzhou 730000, China\\
$^2$School of Nuclear Science and Technology, University of Chinese Academy of Sciences, Beijing 100049, China\\
$^3$School of Nuclear Science and Technology, Lanzhou University, Lanzhou 730000, China}

\begin{abstract}

Based on the Isospin-dependent Boltzmann-Uehling-Uhlenbeck (IBUU) transport model, the in-medium nucleon-nucleon inelastic scattering ( which is dominated
by pion production at low and intermediate energies) is explored. It is found that the in-medium modification of nucleon-nucleon inelastic scatterings appears to reduce the neutron to proton ratio n/p at higher kinetic energies. Although the in-medium modification of nucleon-nucleon inelastic scatterings, as expected, affects the value of $\pi^-/\pi^+$ ratio, considering a series of undetermined properties of delta resonance and $\pi$ in medium, the energetic neutron to proton ratio n/p is more suitable to be used to probe the in-medium correction of nucleon-nucleon inelastic scatterings.

\end{abstract}

\maketitle

\section{Introduction}

Dynamical $\pi$ production in heavy-ion collisions at intermediate energies has been included into transport models more than three decades ago \cite{cugn80,kita86,bertsch88,bauer89,libauer91,danie91,songko2015}. In the models, $\pi$'s are usually treated via resonance decay while the resonance experiences formation and reabsorption in the whole collision process \cite{mosel2015,liba93,liko95,liba2002,xuko2013}. The channels relating to resonance or pion productions are usually implemented in free space because the in-medium correction of nucleon-nucleon inelastic cross section is still undetermined \cite{ditoro2,Lar01,Lar03,Ber88,Mao97,factor,yongepja2010,liq2017,zhangz2017,cui2018}. Uncertainties of the in-medium nucleon-nucleon inelastic cross section in heavy-ion collisions surely affect $\pi^{-}/\pi^{+}$ ratio \cite{yongepja2010,zhangz2017}, an observable which may be used to constrain the high-density symmetry energy \cite{libaprl2002,xiao09,yongjpj2019}. The latter has been considered to play crucial role in many aspects including nuclear physics and astrophysics \cite{topic}.

The reduction of the in-medium elastic cross section of nucleon-nucleon scattering has currently been relatively well determined, see, e.g., Refs. \cite{ligq93,ligq94,alm1995,fuchs2005,factor,liq2006,yongplb11,zhangy2012}.
Although great efforts have been made to find experimental observables constraining the density-dependence of the symmetry energy, little work has been done so far to \emph{probe} the in-medium effects of nucleon or resonance inelastic scattering cross sections, especially in isospin asymmetric nuclear matter.
Because the in-medium reductions of baryon-baryon scattering cross section are not only momentum and density dependent, but isospin-dependent \cite{yongepja2010,factor,liq2017,cui2018}, double ratio of the $\pi^{-}/\pi^{+}$ from neutron-rich and neutron-deficient reactions with the same isotopes thus cannot cancel out uncertainties of the in-medium effects of the baryon-baryon scattering cross section \cite{guowm2014}. What is more, for heavy system, such as Au+Au system, the counterpart system with the same isotopes is not easy to find. Therefore, it is meaningful to find a way to probe and constrain the in-medium baryon-baryon inelastic scattering cross section. Unfortunately, untill now there are rare studies on this subject. $\pi$ production in heavy-ion collisions is thought to be not only affected by the in-medium nucleon-nucleon inelastic scattering cross section but also the \emph{unconstrained} symmetry energy \cite{liyz2005}. The $\pi$ production in heavy-ion collisions is related to a series of undetermined properties of delta resonance and pion in medium (such as delta potential, delta mass and width in medium, pion potential, its momentum, asymmetry and density dependence), which complicates the question. So the straightforward and good way is to use nucleon observable. In heavy-ion collisions at intermediate energies, most energetic nucleons in fact experience the inelastic process, such process may shed interesting light on the exploration of the nucleon-nucleon inelastic scattering cross section in medium. In this study, the in-medium effects of nucleon-nucleon inelastic cross section on the neutron to proton ratio n/p at higher kinetic energies are investigated. It is shown that the energetic n/p ratio is indeed sensitive to the in-medium effects of nucleon-nucleon inelastic cross section. If only the in-medium correction of nucleon-nucleon inelastic cross section is pinned down at several science-based facilities, one could possibly to better constrain the density-dependent symmetry energy by using the $\pi^{-}/\pi^{+}$ ratio \cite{libaprl2002,xiao09,yongjpj2019} as those have been done at FOPI/GSI \cite{fopiw} and S$\pi$RIT/RIBF \cite{spiplb}.

\section{Model description}

In the present study, the Isospin-dependent
Boltzmann-Uehling-Uhlenbeck (IBUU) transport model used includes nucleon-proton short-range correlations, isospin-dependent in-medium elastic and free/in-medium inelastic baryon-baryon cross sections as well as the momentum-dependent isoscalar and isovector pion potentials \cite{yong20151,yong20152,yong20153}. The isospin- and momentum-dependent single nucleon potential reads \cite{spp1,yongpi2017,yongliph2017}
\begin{eqnarray}
U(\rho,\delta,\vec{p},\tau)&=&A_u(x)\frac{\rho_{\tau'}}{\rho_0}+A_l(x)\frac{\rho_{\tau}}{\rho_0}\nonumber\\
& &+B(\frac{\rho}{\rho_0})^{\sigma}(1-x\delta^2)-8x\tau\frac{B}{\sigma+1}\frac{\rho^{\sigma-1}}{\rho_0^\sigma}\delta\rho_{\tau'}\nonumber\\
& &+\frac{2C_{\tau,\tau}}{\rho_0}\int
d^3\,\vec{p^{'}}\frac{f_\tau(\vec{r},\vec{p^{'}})}{1+(\vec{p}-\vec{p^{'}})^2/\Lambda^2}\nonumber\\
& &+\frac{2C_{\tau,\tau'}}{\rho_0}\int
d^3\,\vec{p^{'}}\frac{f_{\tau'}(\vec{r},\vec{p^{'}})}{1+(\vec{p}-\vec{p^{'}})^2/\Lambda^2},
\label{buupotential}
\end{eqnarray}
where $\rho_0$ stands for saturation density, $\tau, \tau'=1/2(-1/2)$ for neutron (proton) and $\delta$ is the isospin asymmetry of the medium. Detailed parameter values can be found in Ref. \cite{yongpi2017}. The symmetry energy's stiffness parameter $x$
is optional and can be used to mimic different forms of the symmetry energy predicted by various
many-body theories without changing any
property of the symmetric nuclear matter and the symmetry energy
at normal density. In this study, we fix x = 1 \cite{yongpi2017} since the effects of the symmetry energy are not the question focused here.

We assume the potential for $\Delta$ resonances is a weighted average of those for neutrons and protons. The weighted factor depending on the charge state of the resonance is the square of the Clebsch-Gordon coefficients for isospin coupling in the processes $\Delta\leftrightarrow \pi N$ \cite{liba2002}, i.e.,
\begin{eqnarray}
\begin{split}
U_{\Delta^-}&=U_{n},\\
U_{\Delta^0}&=\frac{2}{3}U_{n}+\frac{1}{3}U_{p},\\
U_{\Delta^+}&=\frac{1}{3}U_{n}+\frac{2}{3}U_{p},\\
U_{\Delta^{++}}&=U_{p}.
\end{split}
\end{eqnarray}

The nucleon-nucleon free inelastic isospin decomposition cross sections
\begin{equation}
\begin{split}
\sigma^{pp\rightarrow n\Delta^{++}}&=\sigma^{nn\rightarrow p\Delta^{-}}=\sigma_{10}+\frac{1}{2}\sigma_{11},\\
 \sigma^{pp\rightarrow p\Delta^{+}}&=\sigma^{nn\rightarrow n\Delta^{0}}=\frac{3}{2}\sigma_{11},\\
 \sigma^{np\rightarrow p\Delta^{0}}&=\sigma^{np\rightarrow n\Delta^{+}}=\frac{1}{2}\sigma_{11}+\frac{1}{4}\sigma_{10}
\end{split}
\end{equation}
are parameterized via
\begin{equation}
    \sigma_{II'}(\sqrt{s})=\frac{\pi(\hbar c )^{2}  }{2p^{2}}\alpha(\frac{p_{r}}{p_{0}})^{\beta}\frac{m_{0}^{2}\Gamma^{2}(q/q_{0})^{3}}
    {(s^{\ast}-m_{0}^{2})^{2}+m_{0}^{2}\Gamma^{2}}
\end{equation}
with $I$ and $I'$ being the initial state and final state isospins of two nucleons.
The parameters $\alpha, \beta, m_{0}, \Gamma$ as well as other kinematic quantities can be found in Ref.~\cite{VerWest1982}. The cross section for the two-body free inverse reaction is calculated by the modified detailed balance, which takes into account the
finite width of baryon resonance \cite{danie91,liba93}
\begin{equation}
    \sigma_{N\Delta\rightarrow NN}=\frac{m_{\Delta}p_{f}^{2}\sigma_{NN\rightarrow N\Delta}}{2(1+\delta)p_{i}}
    \bigg/\int_{m_{\pi}+m_{N}}^{\sqrt{s}-m_{N}}\frac{dm_{\Delta}}{2\pi}P(m_{\Delta}).
\end{equation}

We extend the effective mass scaling
model for the in-medium elastic cross sections to the inelastic case in isospin
asymmetric matter \cite{factor,yongpi2017}. Compared with the free-space baryon-baryon scattering cross section $\sigma _{free}$, the baryon-baryon scattering cross section in medium $\sigma_{medium}$ is reduced by a factor
\begin{eqnarray}
R_{medium}(\rho,\delta,\vec{p})&\equiv& \sigma
_{medium}/\sigma
_{free}\nonumber\\
&=&(\mu _{BB}^{\ast }/\mu _{BB})^{2},
\end{eqnarray}
where $\mu _{BB}$ and $\mu _{BB}^{\ast }$ are the reduced masses
of the colliding baryon pairs in free and medium cases,
respectively. The effective mass of baryon in medium reads \cite{enefuchs2005}
\begin{equation}
\frac{m_{B}^{\ast }}{m_{B}}=1/(1+\frac{m_{B}}{p}\frac{%
dU}{dp}).
\end{equation}
\begin{figure}
  \includegraphics[width=0.5\textwidth]{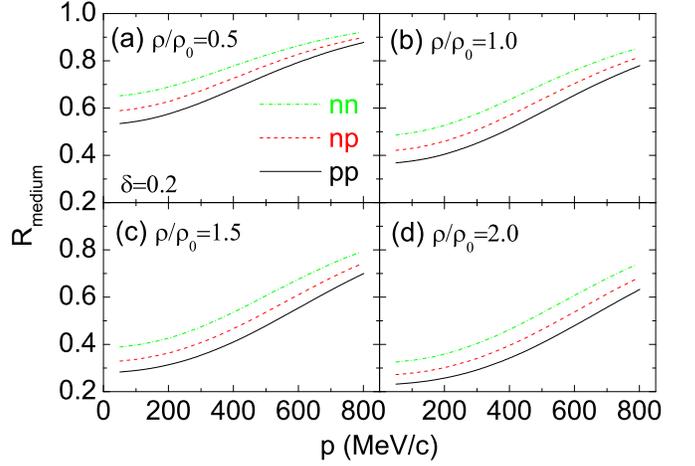}
  \caption{(Color online) Reduction factor $R_{medium}$ of $nn$, $np$ and $pp$ colliding pairs as a function of momentum with different densities ($\rho/\rho_0$=0.5, 1.0, 1.5 and 2.0). The asymmetry of nuclear matter is set to be $\delta$=0.2.}\label{reduce}
\end{figure}

It was argued that the in-medium reduction of the baryon-baryon inelastic cross sections is final-state dependent \cite{gibuu2012,cozma16,liq2017}. The reduction factors connecting with the effective masses of final state particles, e.g., as proposed in Refs. \cite{gibuu2012,cozma16}, are both checked in our model. Compared with our present form used, the $\pi^{-}/\pi^{+}$ ratio corresponding to the reduction factor used in Ref.~\cite{gibuu2012} seems too small. But the $\pi^{-}/\pi^{+}$ ratio corresponding to the form used in Ref.~\cite{cozma16} is quite close to the result of our present model calculation. Based on our single particle potential assumption, it is also found that the above two forms of the reduction factor \cite{gibuu2012,cozma16} are in fact very insensitive to the final-state outgoing channels. So in the present study, for different scattering outgoing channels, we do not consider the splitting of the reduction factor.

Fig.~\ref{reduce} shows the reduction factors of the nucleon-nucleon scattering cross section in medium as a function of nucleon momentum with different densities. Compared with the free-space values, it is seen that the in-medium cross sections are not only reduced, but the reduction factors $R_{medium}$ of $nn$ and $pp$ split. The reduction factor related to neutron is overall larger than that related to proton. To see the in-medium effects of the inelastic cross section on some observables, we in the following make comparative studies of the relevant observables with the free or in-medium baryon-baryon inelastic cross sections.

\section{Results and discussions}

\begin{figure}
  \centering
  \includegraphics[width=0.5\textwidth]{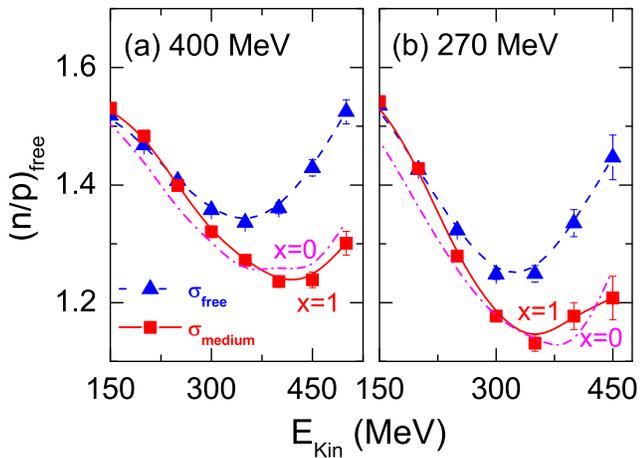}\\
  \caption{(Color online) Effects of the in-medium baryon-baryon inelastic scattering cross section on the n/p ratio in the central collision $^{132}$Sn+$^{124}$Sn at beam energies of 400 (a) and 270 (b) MeV/nucleon. For the in-medium cases, to see the effects of the symmetry energy, the parameter $x$ in Eq.(\ref{buupotential}) is switched from x = 1 (soft) to x = 0 (stiff).}\label{npE400}
\end{figure}
With increase of the baryon-baryon colliding energy, more and more inelastic scatterings occur in heavy-ion collisions. The energetic nucleon spectrum should be affected by the in-medium baryon-baryon inelastic cross sections. Fig.~\ref{npE400} shows the effects of the in-medium inelastic scattering cross section on the n/p ratio in the central collision $^{132}$Sn+$^{124}$Sn at beam energies of 400 and 270 MeV/nucleon, respectively. In the left and right panels of Fig.~\ref{npE400}, since the elastic baryon-baryon scatterings dominate at lower colliding energies, one can see that at lower kinetic energies, effects of the in-medium baryon-baryon inelastic scattering cross section on the neutron to proton ratio n/p are less pronounced. While the n/p ratio of emitted energetic nucleons is clearly affected by the in-medium inelastic baryon-baryon cross section. With increase of the nucleon-nucleon colliding energy, the inelastic scatterings gradually become dominant, it is thus not surprising to see a larger effect of the in-medium inelastic cross section on the energetic neutron to proton ratio n/p. As the reduction of scattering cross section, less nucleons accumulate more energies through multi-scatterings, thus the numbers of the energetic nucleons decrease. Compared with neutrons, positively charged protons suffer more and direct affections from the Coulomb interactions. The latter, to some extend, cancel out the reduction of proton related scatterings in medium. Therefore the decrease of the number of energetic neutron is more evident than that of proton. So the reduction of the in-medium inelastic cross section decreases the value of the energetic neutron to proton ratio n/p. From the both panels of Fig.~\ref{npE400}, it is seen that the effects of the in-medium baryon-baryon inelastic scattering cross section on the energetic n/p ratio reach about measurable 15\%.

For the two incident beam energies of 400 and 270 MeV/nucleon, since we focus on the same nucleon kinetic energy region, i.e., roughly the same 300 -- 450 kinetic energy range, the colliding energies of two nucleons in the two reactions are roughly the same. It is thus not surprising to see almost the same effects of the in-medium reduction of the baryon-baryon inelastic cross section on the energetic n/p ratio.

The effects of the in-medium baryon-baryon elastic scattering cross section on the energetic n/p ratio are in fact also checked. It is found that, with nucleon kinetic energy above 300 MeV, the maximum effects of the in-medium baryon-baryon elastic scattering cross section are just about one half of the effects of the in-medium baryon-baryon inelastic scattering cross section, not mentioning the relatively well determined in-medium baryon-baryon elastic scattering cross section. We therefore expect the experimental measurements of free neutrons and protons with kinetic energies above 300 MeV, by which one may constrain the in-medium reduction of the baryon-baryon inelastic scattering cross section.

It is known that the squeezed-out energetic neutron to proton ratio n/p is very sensitive to the density-dependent symmetry energy \cite{yongplb2007,guoyfprc2019}. To see whether the present analyzed energetic n/p ratio is sensitive to the symmetry energy or not, for the in-medium cases in Fig.~\ref{npE400}, the symmetry energy parameter from x= 1 (soft) to x= 0 (stiff) is switched. One can clearly see that the symmetry energy really plays negligible role in the non-squeezed-out energetic neutron to proton ratio n/p.

\begin{figure}
  \centering
  \includegraphics[width=0.5\textwidth]{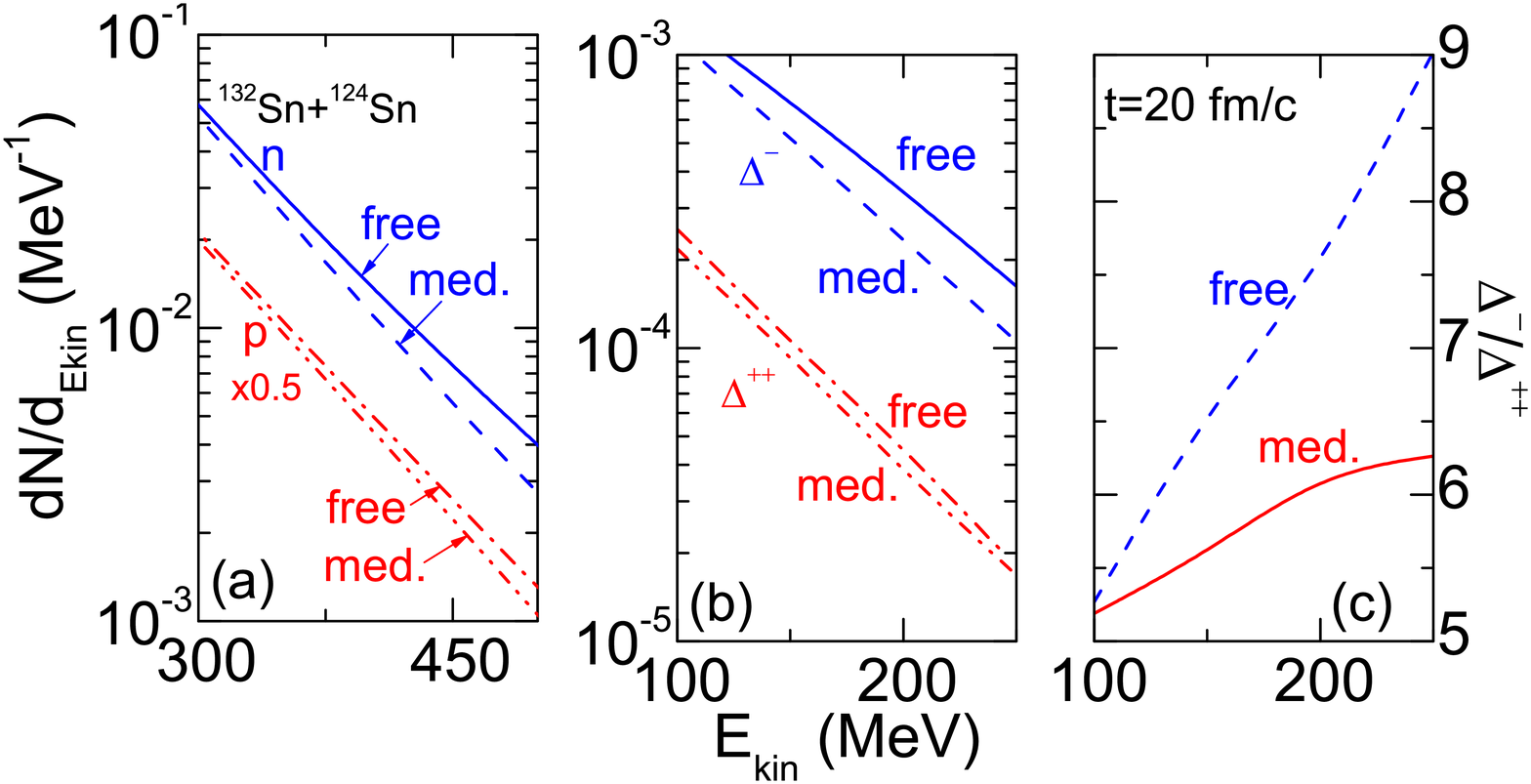}\\
  \caption{(Color online) Effects of the in-medium baryon-baryon inelastic scattering cross section on the energetic nucleons (a), resonances (b) and the ratio of charged resonances at the maximum compression stage (c) in the central collision $^{132}$Sn+$^{124}$Sn at the beam energy of 400 MeV/nucleon.}\label{mE400}
\end{figure}
To further demonstrate the effects of the in-medium baryon-baryon inelastic scattering cross section on particle production, we plot Fig.~\ref{mE400}, energetic nucleons, resonances and the ratio of charged resonances at the maximum compression stage in the central $^{132}$Sn+$^{124}$Sn reactions at 400 MeV/nucleon. As discussed in Fig.~\ref{npE400}, it is seen from the left panel of Fig.~\ref{mE400} that the in-medium baryon-baryon inelastic scattering cross section decreases the energetic nucleons, especially the energetic neutrons. This is the reason why one sees the in-medium baryon-baryon inelastic scattering cross section reduces the ratio of energetic neutron to proton ratio n/p as shown in Fig.~\ref{npE400}. Since the energetic nucleon-nucleon collision produces $\Delta$ resonance and the proton-proton (neutron-neutron) collision produces $\Delta^{++}$ ($\Delta^{-}$), one sees similar effects of the in-medium baryon-baryon inelastic scattering cross section on the energetic resonance productions as shown in the middle panel of Fig.~\ref{mE400}. In the right panel of Fig.~\ref{mE400}, the effects of the in-medium baryon-baryon inelastic scattering cross section on the ratio of $\Delta^{-}/\Delta^{++}$ are clearly shown. Because the productions of the energetic $\Delta^{-}$ or $\Delta^{++}$ directly connect with the energetic neutron-neutron or proton-proton collisions and the number of the energetic neutrons reduces more than that of the energetic protons, the reduction of the in-medium baryon-baryon inelastic scattering cross section thus decreases the ratio of the $\Delta^{-}/\Delta^{++}$. The energetic nucleons studied here are not the result of iso-fractionation, but the result of nucleon-nucleon scatterings. So the energetic neutron to proton ratio n/p is a direct reflection of the energetic nucleon isospin components of the dense matter formed in heavy-ion collisions, rather than the other way round.

For energetic nucleon collisions, the produced $\Delta$ resonances possess relatively low kinetic energies. Also at any time, $\Delta$ resonances experience decays and re-absorptions. The difference of the number of the very high energy nucleons and the number of primordial produced low energy $\Delta$ resonances is in fact not so large, thus the $\Delta$ resonances in the inelastic collisions influence the dynamics of the collisions of the very high energy nucleons. At 400 -- 500 MeV per nucleon kinetic energy (the colliding pair has totally 800 -- 1000 MeV kinetic energy), the value of the inelastic cross section is in fact larger than the value of the elastic cross section, effects of the inelastic cross section are thus larger than that of the elastic cross section. In this case, the modification of the inelastic cross section evidently affects the dynamics of the collision of the energetic nucleons.

\begin{figure}
  \centering
  \includegraphics[width=0.5\textwidth]{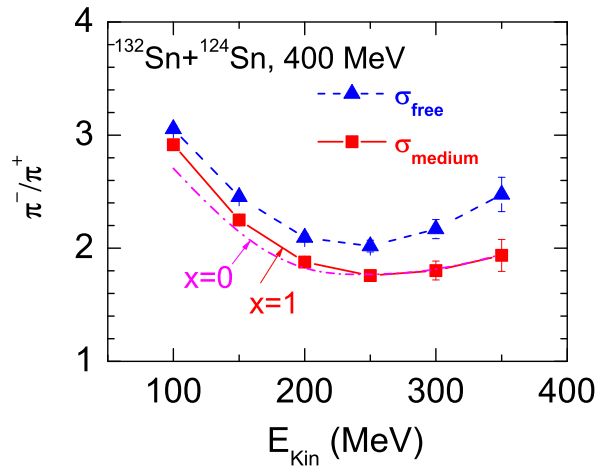}\\
  \caption{(Color online) Effects of the in-medium baryon-baryon inelastic scattering cross section on the kinetic energy distribution of the $\pi^-/\pi^+$ ratio in the central collision $^{132}$Sn+$^{124}$Sn at the beam energy of 400 MeV/nucleon. For the in-medium case, to see the effects of the symmetry energy, the parameter $x$ in Eq.(\ref{buupotential}) is switched from x = 1 (soft) to x = 0 (stiff).}\label{piE400}
\end{figure}
Since $\Delta^{++}$ ($\Delta^{-}$) decays into $\pi^{+}$ ($\pi^{-}$), it is expected that the reduction of the inelastic baryon-baryon scattering cross section in medium also decreases the ratio of the energetic $\pi^{-}/\pi^{+}$ ratio. Fig.~\ref{piE400} shows the kinetic energy distribution of the energetic $\pi^-/\pi^+$ ratio in the central reaction $^{132}$Sn+$^{124}$Sn at 400 MeV/nucleon incident beam energy with free or in-medium baryon-baryon inelastic scattering cross sections. It is seen that the ratio of the energetic $\pi^{-}/\pi^{+}$ ratio reduces about 15\% when the in-medium inelastic baryon-baryon scattering cross section is employed.

In inelastic process with isospin-dependent potential, collision threshold modification related to resonance and pion production is not considered here \cite{zhangz2017,cozma16}. The reason is that an inelastic process is a high energy process relative to the potential difference of ingoing and outgoing channels. If one considers an inelastic process as an instant process and the potential difference can be remedied by surrounding matter, one may neglect the collision threshold modification since for all the inelastic scattering processes the total energy is conserved.

Although the $\pi^{-}/\pi^{+}$ ratio is frequently mentioned while probing the high-density symmetry energy, its kinetic distribution at high energies is not sensitive to the symmetry energy. This result in fact does not conflict with that shown in Ref.~\cite{gaoyuan2013}. The latter is for squeezed-out pion emission whereas our present analysis is for general pion emission. As mentioned before, pion production in heavy-ion collisions relates to a series of undetermined properties of delta resonance (such as delta potential, delta mass and width in medium) and pion in medium (such as pion potential, its momentum, asymmetry and density dependence), thus it is not suitable to probe the in-medium baryon-baryon inelastic scattering cross section by pion production.

\section{Conclusions}

In summary, based on the transport model IBUU, effects of the in-medium baryon-baryon inelastic scattering cross section on the energetic nucleons, delta resonances and pions are studied.
It is found that in $^{132}$Sn+$^{124}$Sn reaction at intermediate energies the in-medium baryon-baryon inelastic scattering cross section reduces the emitting numbers of energetic nucleons, especially energetic neutrons. Therefore the in-medium baryon-baryon inelastic scattering cross section decreases the energetic neutron to proton ratio n/p. As the energetic nucleons in the reaction directly relate to the  energetic delta resonance productions, the resonance yields and their ratio are also sensitive to the in-medium baryon-baryon inelastic scattering cross section. Because pion production is closely connected with the delta resonance, the ratio of $\pi^{-}/\pi^{+}$ is expectedly sensitive to the in-medium baryon-baryon inelastic scattering cross section too. In the study, it is also seen that, due to the non-squeezed-out emissions, both the energetic neutron to proton ratio n/p and the energetic charged pion ratio $\pi^{-}/\pi^{+}$ are not sensitive to the symmetry energy. Since the pion production in heavy-ion collisions relates to a series of undetermined properties of delta and pion in medium, the energetic neutron to proton ratio n/p is more suitable than the energetic $\pi^{-}/\pi^{+}$ ratio to be used to probe the in-medium baryon-baryon inelastic scattering cross section. The present studies can be conducted at, such as, the GSI Facility for Antiproton and Ion Research (FAIR) in Germany \cite{fopi16}, the Cooling Storage Ring on the Heavy Ion Research Facility at IMP (HIRFL-CSR) in China \cite{csr}, the Facility for Rare Isotope Beams (FRIB) in the Untied States \cite{frib}, the Radioactive Isotope Beam Facility (RIBF) at RIKEN in Japan \cite{sep,shan15,ribf} and the Rare Isotope Science Project (RISP) in Korea \cite{korea}. After this, it then could be possible to constrain the density-dependent symmetry energy by using the $\pi^{-}/\pi^{+}$ ratio \cite{libaprl2002,xiao09,yongjpj2019} because the double ratio of the $\pi^{-}/\pi^{+}$ from the neutron-rich and neutron-deficient reactions with the same isotopes only cancels out a fraction of uncertainties of the in-medium effects of the baryon-baryon inelastic scatterings \cite{guowm2014}.


This work is supported in part by the National Natural Science Foundation of China under Grant Nos. 11775275, 11435014.

\end{document}